\documentclass[article]{revtex4}
\usepackage{pdfpages}
\usepackage{epsf}
\usepackage{subfigure}
\usepackage{amsmath}
\usepackage{eqnarray,amsmath}
\usepackage{epstopdf}
\usepackage{mathrsfs}
\usepackage{natbib}
\usepackage{amssymb,latexsym}
\usepackage{dcolumn}
\usepackage{graphicx}
\usepackage{color}

\makeatletter
\newcommand*{\rom}[1]{\expandafter\@slowromancap\romannumeral #1@}
\makeatother
\def\be{\begin{equation}}
    \def\ee{\end{equation}}
\def\ba{\begin{eqnarray}}
    \def\ea{\end{eqnarray}}

\graphicspath{{images/}}
\begin{document}
    \title{\large \bf Revisiting nonrelativistic limits of field theory: Antiparticles}

       \author{Javad T. Firouzjaee}
        \affiliation{ School of Astronomy, Institute for Research in Fundamental Sciences (IPM), P. O. Box 19395-5531, Tehran, Iran } \email{j.taghizadeh.f@ipm.ir} 
        
        \author{Abasalt Rostami} 
        \affiliation{Department of Physics, Sharif University of Technology, Tehran, Iran } 
        \affiliation{ School of Physics, Institute for Research in Fundamental Sciences (IPM), P. O. Box 19395-5531, Tehran, Iran } 
        \email{aba-rostami@ipm.ir} 
        
        \author{Mahmud Bahmanabadi} 
        \affiliation{Department of Physics, Sharif University of Technology, Azadi Avenue, Tehran, Iran } 
        \email{bahmanabadi@sharif.edu}

    \begin{abstract}
        Different subtleties and problems associated with a nonrelativistic
         limit of the field theory to the Schroedinger theory are discussed. In this paper, we revisit different cases of the nonrelativistic limit of a real and complex scalar field in the level of the Lagrangian and the equation of motion. We develop the nonrelativistic limit of the Dirac equation and action in the way that the nonrelativistic limit of  spin-$\frac{1}{2}$ wave functions of particles and antiparticles appear simultaneously.
         We study the effect of a potential like $U(\phi)\propto \phi^4$ which can be attributed to axion dark matter field in this limit. We develop a formalism for studying the nonrelativistic limit of antiparticles in the quantum mechanics. We discussed the non-local approach for the nonrelativistic limit and its problems.

    \end{abstract}
    %
    %
    \maketitle
    \tableofcontents
    
    \newpage
    
    \section{Introduction}

The relativistic quantum field theory (QFT) is built in the way that the quantum physics becomes Lorentz invariance. The Lorentz invariance grantees that our theory is consistent with (special) relativity.  It is generally accepted that the concept of antiparticles can appear in the context of QFT. This comes from the fact that it is only in this regime that one can have free particles with negative energies traveling backward in time, whose absence is interpreted as positive energy and opposite charge and momentum particles (antiparticles) traveling forwards in time. One can say that we need antiparticle to ensure causality in a Lorentz invariant theory \cite{Pad-book}. But we don't need a relativist theory to describe all physics around us. For example, in Condensed Matter Field Theory, one should not be essentially very concerned about relativistic models, because typical condensed matter systems have thermal energies which are very small in comparison with the energy scale of relativistic theories.\\

Recent advances in science and technology, provide the opportunity to detect the nonrelativistic antiparticle. For example, complete atoms of antimatter have been assembled out of antiprotons and positrons, collected in electromagnetic traps. Moreover, Satellite experiments have established evidence of positrons and a few antiprotons in primary cosmic rays, amounting to less than 1\% of the particles in primary cosmic rays. Since the observation of antiparticles (antiprotons) in cosmic rays \cite{antiparticles-observation}, many studies of cosmic ray antiparticles have been performed which bring motivation to study its nonrelativistic behavior \cite{antiparticles-observation-study}. \\

A similar thing happens in late time universe. The late time matter (e.g
dark matter) in our universe which comes from the quantum fluctuation in the early universe follows nonrelativistic physics in late time universe when the universe becomes cold due to the expansion. It is accepted that many observational evidences indicate that dark matter should be plentiful throughout the universe, contributing roughly thirty percent of all matter (energy) and five times more to the energy density of the universe than baryonic matter. The dark matter should be cold and collisionless. Several candidates for the dark matter particle have been proposed, including weakly interacting massive particles, sterile neutrinos, and axions, among others.  We are interested to study the axion as a scalar field for the nonrelativistic problem which for small field values the axion potential get to  $ \lambda \phi^4$ theory \cite{Guth:2014hsa}.\\
 
 In this paper, we examine nonrelativistic limit in different cases for a scalar field in the level of the Lagrangian and the equation of motion. We develop the nonrelativistic analysis for antiparticles and consider their quantum mechanics.  As an important example of the scalar field potential, we study the $ \lambda \phi^4$ theory which can be attributed to the axion dark matter field in the small field value. Furthermore, we discuss the nonlocal nonrelativistic limit analysis which presented recently by Namjoo,  Guth, and Kaiser (NGK) \cite{Namjoo:2017nia}.  In the last part of the paper, we extend our discussion to the nonrelativistic limit of a Dirac field.\\
 
This paper is organized as follows: In Section II we introduce the nonrelativistic field theory of a complex scalar field. In Section III we study the nonrelativistic limit for real scalar field theory. Section IV is devoted to develop the nonrelativistic limit of the antiparticles and its quantum mechanics. In section V we discuss the nonlocal approach for the nonrelativistic limit. Section VI is devoted to study the nonrelativistic limit of a Dirac field.
Concluding remarks follow in Section VII. In the paper, we keep the speed of light, $c$,  explicitly in the equation.\\

\section{Nonrelativistic limit of a complex scalar field}

It is known that one straight forward way to get the quantum filed equation for a scalar field is to put the operator equation for $E=i \frac{\partial}{\partial t}$ and $p= -i \nabla$ in the following Lorentz invariace equation
\be
E^2 \phi=(P^2 c^2+m^2 c^4) \phi
\ee
in the special relativity to get the Klein-Gordon equation. This equation describes a free scalar field $ \square \phi+m^2 c^2 \phi=0$. The Lagrangian for this complex scalar field theory is
\be
\mathcal{L}=\frac{1}{2}\partial_\mu \phi \partial^\mu \phi^*-\frac{m^2 c^2}{2} \phi \phi^*.
\ee
where $\partial_{\mu}= (\frac{\partial}{c \partial t},\frac{\partial}{ \partial x^{i}})$ and $\partial^{\mu}= (\frac{\partial}{c \partial t},-\frac{\partial}{ \partial x^{i}})$ . The solution for this free complex scalar  can be written as
\be \label{phi-def}
\phi(x)=\int \frac{d^3 p}{(2\pi)^3 \sqrt{2 E_p}} \left( a_p e^{-ipx}+ b_p^\dagger e^{ipx}  \right)=A(x)+ B^\dagger(x)
\ee
where $px=Et-\vec{p}.\vec{x}$. One can get the the Hamiltonian density for this Lagrangian by defining the conjugate momentum $\pi(x)=\frac{\partial \mathcal{L}}{\partial \phi(x)}$. Following the Poisson bracket in the classical mechanics, we (canonically) quantize the field by imposing the following commutation relation at a constant time 
\ba
[ \phi(x),\phi(y)]=[\pi(x), \pi(y)]=0 \\ \nonumber 
[ \phi(x),\pi(y)]=(2\pi)^3\delta^{(3)}(\vec{x}-\vec{y})
\ea
The complex scalar field has an internal U(1) symmetry and one can shows that the Noether current 

\be
j^\mu(x)=\frac{i}{2m}(\phi^* \partial^\mu \phi-\phi \partial^\mu\phi^* )
\ee
satisfy the conservation equation $\partial_\mu j^\mu =0$. In a coordinate base we get 
\be 
\frac{d \rho}{dt} + \nabla.\vec{j}=0
\ee
where $\rho=\frac{i}{2m c^2} (\phi^* \partial_t \phi-\phi \partial_t \phi^*) $ and $ \vec{j}=-\frac{i}{2m}(\phi^* \nabla \phi-\phi \nabla \phi^*) $. It can be  seen from the continuity equation that if the spacial current become zero at far distance or in the boundary of a surface the charge $Q=\int \rho d^3x$ is a constant.\\
In the case that we have potential other than mass term, $V(\phi) = \frac{m^2 c^2}{2} \phi \phi^* + U(\phi,\phi^*)$, the Lagrangian density will be
\be
\mathcal{L}=\frac{1}{2}\partial_\mu \phi \partial^\mu \phi^*-\frac{m^2 c^2}{2} \phi. \phi^*-U(\phi,\phi^*)
\ee
In this case, the Klein-Gordon equation gets $\square \phi+(m^2 c^2)\phi + \frac{dU(\phi,\phi^*)}{d\phi^*} =0$.
In this paper, we consider our example with known $U(\phi,\phi^*)=\frac{\lambda}{4!} (\phi \phi^*)^2$ potential (which is the axion potential at small field value).\\

\subsection{Nonrelativistic limit}

The nonrelativist limit of a field means that the total energy of a field minus the rest mass energy $mc^2$ must be small relative to its mass i.e. $E-mc^2 \ll mc^2$. Mathematically, we perform the (nonrelativistic) transformation to subtract the rest mass energy
\be \label{trans-com}
\phi=e^{-imc^2 t } \psi.
\ee
In other words, the kinetic energy must be small 
\be
(E-mc^2)\phi \approx -i e^{-mc^2t } \partial_t\psi \ll e^{-imc^2t } mc^2 \psi.
\ee
Sometime the nonrelativistic limit is interpreted as $|p|/m=v/c\ll 1$, or equivalently we take the limit
$c\rightarrow \infty$.
In the nonrelativistic domain, the excitation energies of particles are small compared to the particle mass contribution $mc^2$.
Now, lets look at the Lagrangian density in this limit. Due to the transformation Eq. \eqref{trans-com} we have
\be
\partial_t\phi = (\partial_t\psi -imc^2  \psi)e^{-imc^2 t } \approx-imc^2  \psi e^{-imc^2 t }.
\ee
In this case, the Lagrangian density at the leading order gets
\be
\mathcal{L}=\frac{im}{2}(\psi^* \partial_t\psi - \psi \partial_t\psi^*)-\frac{1}{2}\nabla\psi \nabla\psi^*
\ee 
where this is the Schroedinger Lagrangian and the variation relative to the  $\psi^*$ will give the Schroedinger  equation, $i \partial_t\psi=-\frac{\nabla^2}{2m} \psi$.\\ 

There are some points that we have to mention. First, the mass term in the relativistic Lagrangian density is completely removed due to the nonrelativistic transformation without any approximation. Second, the next order relative to the $\frac{1}{c}$ will gives the exact Klein-Gordon action. In other words, there is \textbf{no relativistic correction} in this case. Third, since the potential is real in the Lagrangian density the potential term $U(\phi, \phi^*)$ remain unchanged in Lagrangian due to the nonrelativistic transformation, and the Schroedinger equation reduces to
\be
i  \partial_t\psi=-\frac{\nabla^2}{2m} \psi + \frac{dU(\psi,\psi^*)}{d\psi^*} .
\ee
This equation shows that for a potential which is not proportional to $U(\phi, \phi^*)=v(x) \phi \phi^*$, the Schroedinger equation becomes nonlinear.\\
One can study the nonrelativistic limit in the level of the equation of motion. If we take the second derivative of the scalar field in the nonrelativistic transformation, we get,
\be 
\ddot{\phi}=(\ddot{\psi}-imc^2 
\dot{\psi}-m^2c^4\psi)e^{-imc^2 t}
\ee
If we put this term in the equation of motion, we get the same Schroedinger equation as we get from the Lagrangian. Similar to the Lagrangian there is no relativistic correction for the equation of motion and the next order is Klein-Gordon equation.  In the case of the complex scalar field, the nonrelativistic limit of the Lagrangian and the equation of motion lead to the same results, but in the case of the real scalar field the are some subtleties that we consider in the next section.\\

Using the nonrelativistic transformation, the Noether current and continuity equation will give 
\be
\frac{d(\rho=\psi \psi^*)}{dt}+\nabla.\vec{j}=0
\ee
This is exactly the continuity equation for the quantum mechanics. The key point in the limit is that the mass or charge density, $\rho$, in quantum mechanics is always positive, but in the quantum field theory it can be positive or negative and it cannot be interpreted as a probability density. \\

\section{Nonrelativistic limit of a real scalar field}

The main application of field theory in cosmology and quantum field in curved spacetime is devoted to study the real scalar field. The Lagrangian density for a real scalar field is 
\be
\mathcal{L}=\frac{1}{2}\partial_\mu \phi \partial^\mu \phi-\frac{m^2 c^2}{2} \phi^2-U(\phi)
\ee
To study the nonrelativistic limit, we apply the transformation
\be
\phi=e^{-imc^2 t } \psi + e^{imc^2 t } \psi^*.
\ee
If one put this transformation into the Lagrangian, there are some terms that rapidly oscillate when we take nonrelativistic limit $c\rightarrow \infty$ \footnote{The precise interpretation of the nonrelativistic limit in the exponential means that the $ mc^2\gg \frac{1}{\Delta T }$ where the $\Delta T$ the time scale in which the system is observed}. As discussed in the Appendix A, since these terms are integrated by $t$ in the action, one can average out this terms which appear as  $\mathcal{O}(\frac{1}{c^n})$ where $n>2$ (this is because we usually assume that the time derivative of field or at least its value become zero when one goes to future or past infinity. Otherwise, we can not get the Schroedinger equation in the nonrelativistic limit). In this limit, the Lagrangian density in the leading order for a free scalar field appears
\be
\mathcal{L}=\frac{im}{2}(\psi^* \partial_t\psi - \psi \partial_t\psi^*)-\frac{1}{2}\nabla\psi \nabla\psi^*
\ee
which is the Schroedinger equation action. This Lagrangian has the \textbf{relativistic correction} generally. 

One can take the nonrelativistic limit of the free scalar field in the level of the equation of motion, where in this case the Klein-Gordon equation gets the the following equation for the $\psi$ and $\psi^*$
\be \label{schro-banergee}
e^{-imc^2 t }(2i m \partial_t+\nabla^2-\frac{1}{c^2}\partial_t^2) \psi + e^{imc^2 t } (-2i m \partial_t+\nabla^2-\frac{1}{c^2}\partial_t^2) \psi^*=0
\ee
Authors in \cite{Banerjee:2018pvs} claim that going to the $c \rightarrow \infty$ limit and interpreting $\psi$ and $\psi^*$ as independent fields yields
the pair of equations, $(2i m \partial_t+\nabla^2-\frac{1}{c^2}\partial_t^2) \psi=0$ and  $(-2i m \partial_t+\nabla^2-\frac{1}{c^2}\partial_t^2) \psi^* =0$. The point is that the Eq. \eqref{schro-banergee} only say that the real part of the $\Re\left(  e^{-imc^2 t }(2i m \partial_t+\nabla^2-\frac{1}{c^2}\partial_t^2) \psi\right) =0$ not all term is zero. As a results, we cannot get the Schroedinger equation from the equation of motion. In the case that we have a potential in the Klein-Gordon equation, the oscillating and crossing term $\psi$ and $\psi^*$ make problems to get the Schroedinger equation.

In the case of the power law potential like $U(\phi)=\frac{\lambda}{4!} \phi^4$, the oscillating terms are disappeared in the action and we get known the Lagrangian density
\be \label{sch-lag-real}
\mathcal{L}=\frac{im}{2}(\psi^* \partial_t\psi - \psi \partial_t\psi^*)-\frac{1}{2}\nabla\psi \nabla\psi^*-\frac{\lambda}{16m^2} \psi \psi^*
\ee
It can be seen that the oscillating terms remain in the potential if one reduces the Klein-Gordon equation in the nonrelativistic limit. This is because we could not average out these terms in the equation of motion as we did in the action. Since we start studying a theory in the action level, not in the equation of motion, therefore we can not trust these oscillation terms in the equation of motion. \\

The real scalar field does not have mass or charge density and its four current vanishes, $j^\mu=0$. But since the reduced Lagrangian density Eq. \eqref{sch-lag-real} has $U(1)$ symmetry, it has conserved charge or mass density.\\

\section{Antiparticles quantum mechanics}

In this section we do similar analysis for antiparticles as did in the reference \cite{Pad-book}. After quantizing Eq. \eqref{phi-def}, in the field theory \textit{one antiparticle state} is built $<0|\phi^\dagger(x)$. This quantity is not a c-number to study its nonrelativistic behavior. One can define antiparticle wave function as
\be
\bar{\psi}(t,x)=<0|\phi^\dagger(t,x)|\psi>=<0|B(x)|\psi>,
\ee
where $|\psi>$ in an arbitrary state. Taking the time derivation, we get
\be
i\partial_t\bar{\psi}(t,x)= i\partial_t<0|\phi^\dagger(t,x)|\psi>
=<0|\int \frac{d^3 p~~\sqrt{p^2c^2+m^2c^4}}{(2\pi)^3  \sqrt{2 E_p}} \left( b_p e^{-ipx} - a_p^\dagger e^{ipx}  \right)|\psi>.
\ee
After some calculation and using $p=-i \nabla$, one gets
\be
i\partial_t\bar{\psi}(t,x)=<0|\sqrt{m^2c^4- c^2 \nabla^2} \phi^\dagger(t,x)|\psi>.
\ee
Taking nonrelativistic limit with $c\gg 1$ we get
\be
i\partial_t\bar{\psi}(t,x)=(mc^2-\frac{1}{2m}\nabla^2+\mathcal{O}(\frac{1}{c^2}))\bar{\psi}(t,x).
\ee

The first term on the right hand side is the rest energy  which will be
removed by the nonrelativistic transformation $\bar{\psi}(t,x)=e^{-imc^2 t } \bar{\varphi}(t,x)$. Thus, we get the Schroedinger equation for antiparticle wavefunction
\be
i
\partial_t \bar{\varphi}(t,x)=-\frac{1}{2m}\nabla^2 \bar{\varphi}(t,x).
\ee

Since in this method we used field extension in terms of creation and annihilation operator, it is not possible to extend this method for general potential which we do not know his creation and annihilation operator extension.\\

Another way to get the Schroedinger equation for the antiparticles \cite{Padmanabhan:2017bll} is to expand the action in term of $\phi(x)=a(x)+b^\dagger(x)$ and define nonrelativistic transformation for particle and antiparticle as
\be
a(x)=e^{-mc^2 t}<0|A(x)|\psi>,~~b(x)=e^{-mc^2 t}<0|B(x)|\psi>. 
\ee
It can be seen that if one average out the oscillating term (see Appendix A for more detail), the Klein-Gordon action gets the Schroedinger action for the particle and antiparticle. This shows antiparticles do not go away when we take the nonrelativistic limit for a complex scalar field. This procedure works even for a real scalar field for which the antiparticle is identical to
the particle. In this nonrelativistic case,
particles and antiparticles are conserved separately. \\

\section{Non-local approach for nonrelativistic limit}

In this section we discuss what NGK did in \cite{Namjoo:2017nia} to extend the nonrelativistic analysis for a real scalar field. The authors in \cite{Namjoo:2017nia} aimed to obtain an expression for the Lagrangian that yields the Schroedinger equation as the effective equation of motion for the redefined field in the nonrelativistic limit. Here, we will first review
what they did and then explain the issues of their work.
The authors start with common Lagrangian density for a real scalar field with a self interaction of forth order:

\be \label{e1}
{\cal L} = \frac{1}{2} \eta^{\mu\nu} \partial_\mu \phi \partial_\nu \phi - \frac{1}{2} m^2 c^2 \phi^2 - \frac{1}{4!} \lambda \phi^4 .
\ee

The equation of motion for such system would be given as,

\ba \label{e2}
\begin{split}
    \frac{1}{c}\dot \phi &= {\delta H \over \delta \pi} = \pi ,\\
   \frac{1}{c} \dot \pi &= - {\delta  H \over \partial \phi}  = (\nabla^2 - m^2 c^2) \phi - \frac{1}{3!} \lambda \phi^3 .
\end{split}
\ea

Here the Hamiltonian, $H$, is defined as a spatial integration over the density Hamiltonian 

\be \label{e3}
{\cal H} = \frac{1}{2c^2} \pi^2 + \frac{1}{2} ({\bf \nabla} \phi)^2 + \frac{1}{2} m^2 c^2 \phi^2 + \frac{1}{4!} \lambda \phi^4  .
\ee

If one would like to derive the nonrelativistic limit of the equation of motion  or Lagrangian, it is enough to define a new field $\psi$ as following 

\ba \label{e4}
\phi (t, {\bf x} )&= \frac{1}{\sqrt{2mc^2} } \left[ e^{-imc^2t} \psi (t, {\bf x}) + e^{imc^2t} \psi^*  (t, {\bf x} ) \right] .
\ea

With the help of this definition, one can easily show that in the limit $c \rightarrow \infty$, the Klein-Gordon action reduces to the Schroedinger action with the Lagrangian density 

\be \label{e5}
{\cal L} = \frac{i}{2} \left( \dot{\psi} \psi^* - \psi \dot{\psi}^* \right) - \frac{1}{2m} \nabla \psi \nabla \psi^* .
\ee

At this point the authors introduce a momentum density in the nonrelativistic limit as
\ba \label{e6}
\pi (t, {\bf x} ) &= - i \sqrt{ \frac{ mc}{2} } \, \left[ e^{-imc^2t} \psi (t, {\bf x}) - e^{imc^2t} \psi^*  (t, {\bf x} ) \right] .
\ea

They claim that plugging Eqs. (\ref{e6}, \ref{e4}) into the Eqs. (\ref{e2}) and neglecting oscillating terms, one reaches to the usual Schroedinger equation. At the next step, to reach a more different sense of nonrelativistic limit, they try to use a nonlocal redefinition of $\phi$ as 

\ba \label{e7}
\begin{split}
    \phi (t, {\bf x}) &= \frac{1}{ \sqrt{ 2mc} } {\cal P}^{-1/2} \left[ e^{-imc^2t} \psi (t, {\bf x} ) + e^{imc^2t} \psi^* (t, {\bf x} ) \right] , \\
    \pi (t, {\bf x} ) &= - i \sqrt{ \frac{ mc}{2} } \, {\cal P}^{1/2} \left[ e^{-imc^2t} \psi (t, {\bf x} ) - e^{imc^2t} \psi^* (t, {\bf x} ) \right] ,
\end{split}
\ea

where ${\cal P } \equiv \sqrt{ 1 - \frac{ \nabla^2}{m^2 c^2} }$ \footnote{From the equation of motion we know that $\pi = \dot{\phi}$, but one can check that the equation (7) in the NGK paper does not satisfy this equation generally and we have to make a constraint on $\psi$ to satisfy this equation. If one considers the momentum properly in the equation (7) with nonlocal field redefinition which satisfies $\pi = \dot{\phi}$, one lost the commutation relation in the equation (10).}. Using the equation of motion  one finds

\be \label{e8}
i \dot{\psi} = mc^2 \left( {\cal P} - 1 \right) \psi + \frac{ \lambda e^{imc^2t} }{4! \, m^2 } \, {\cal P}^{-1/2} \left[ e^{-imc^2t} \, {\cal P}^{-1/2} \psi + e^{imc^2t} \, {\cal P}^{-1/2} \psi^* \right]^3 .
\ee

In the case of $\lambda=0$ it is easy to show that the above equation reduces to the free Schroedinger equation when one takes the limit $c\rightarrow \infty $. There are some issues when one studies Eq. (\ref{e8}). First, we suppose $\lambda=0$. In this case, one easily finds (apart from the phase $e^{imt}$) the famous differential equation which people had found in the first attempt to drive a relativistic version of quantum mechanics. Although they had found a relativistic equation, because of some reasons they threw it out; There is a first order in the time derivative but infinite-order in space derivatives which requires having the infinite number of initial conditions.   Another problem is that because of nonlocality, this equation can violate causality. Also, the operator $\cal{P}$ which appears here is not well-defined and in fact has a singular kernel. These problems remain even when one tries to use quantum field theory perspective. One might understand why such problems arise when he or she tries to find a solution for this kind of differential equations. In fact $i \dot{\psi} = {\cal P}  \psi$ can have a solution like

\ba \label{ee9}
\psi(t,\vec{x})=\int d^3 p ~ f(p) \exp^{-ip.x},
\ea
where $px=p^0t - \vec{p}.\vec{x}$ and $p^0 = +\sqrt{\vec{p}^2 + m^2 c^2}$. Therefore, here we have only the positive frequency mode and there is not any negative frequency mode sector in the solution.  One can see that this  equation $i \dot{\psi} = m \left( {\cal P} - 1 \right) \psi $ has not all information of the field $\phi$. This equation is first order in time derivative and one can have the $\psi(t,\vec{x}) $ solution in all time if one knows it in the initial time $\psi(t_0,\vec{x})$ but in Klein-Gordon equation we need both $\phi(t,\vec{x}) $ and $\dot{\phi}(t,\vec{x}) $  to specify it at all time. This is because we lost negative frequency mode with choosing this type of the nonlocal redefinition \eqref{e7}. The negative frequency mode has a crucial role in the concept of causality and without this sector we lave lost the causality. \\

Now, lets come back to the field redefinition in the Eqs. (\ref{e7}). This kind of field redefinitions have big problems: They remove all negative frequency modes and so shouldn't be regarded as physical theories. Moreover, we have to mention an important note about the interaction term here; If we assume the field redefinition in Eq. (\ref{e4}) and write down the equation of motion in the presence of interaction part $\lambda \phi^3$, then the nonrelativistic limit of the equation of motion would be

\be \label{ee8}
i \dot{\psi} = -\frac{\nabla^2}{2m}\psi +\frac{\lambda}{8m^2}|\psi|^2\psi.
\ee

To reach this equation one has to neglect all fast-oscillating terms without concerning about their backreactions. More precisely, this happens because the backreaction of the positive frequency modes would be canceled out with the negative frequency ones. Therefore, we expect the interaction part in the Eq. (\ref{e8}) includes extra backreaction effects (because of neglecting the frequency modes) which prohibit to neglect fast-oscillating terms.

One may think that the effect of the negative frequency modes can be negligible in the nonrelativistic limit but in fact, it is not true. People have shown that the correct reduction of relativistic quantum field theory to the nonrelativistic one has to include the effect of the negative frequency modes \cite{Padmanabhan:2017bll}.\\

\section{Nonrelativistic limit of a fermion}

In this section, we try to find a nonrelativistic action for the Dirac field. Having found the nonrelativistic Dirac action, one can use it to write down the corresponding green function of such theory using Feynman path integral.

\subsection{One particle state of a fermion}

In particle physics, the Dirac equation describes all spin-$\frac{1}{2}$ massive particles such as electrons and quarks for which parity is a symmetry.
One can write the Dirac free field operator a
\be \label{Di-phi-def}
\phi(x)=\int \frac{d^3 p}{(2\pi)^3 \sqrt{2 E_p}} \sum_{s} \left( a_p^s u^s(p) e^{-ipx}+ b_p^{s \dagger} v^s(p) e^{ipx}  \right)
\ee

where the operators $ a_p^{s\dagger} $ create particles associated to the spinors $ u^s(p) $ while $ b_p^{s\dagger} $ create particles associate to $ v^s(p) $.  The $ s $ index is the 2-component spin index of fields. Similar to the scalar field case one can construct one particle state of a fermion field  as $<0|\phi(x)$ and  its antiparticle field, $<0|\phi^\dagger (x)$ respectively.\\

\subsection{Nonrelativistic limit of Dirac action}

The Lagrangian density for a Dirac field is given by
\be
\mathcal{L}=\bar{\psi}(x)(i\gamma^\mu \partial_\mu-m)\psi(x).
\ee
where $\gamma^0=\begin{bmatrix}
1 & 0  \\
0 & -1  \\
\end{bmatrix}$ and $\gamma^i=\begin{bmatrix}
0 & \sigma^i  \\
-\sigma^i  & 0  \\
\end{bmatrix}$  are  Dirac matrices and $\sigma^i$ present Pauli matrices. One can decompose the spinor in terms of two two-component objects, $\psi=\begin{pmatrix}
 \psi_L\\
\psi_R
\end{pmatrix}$. Taking the variation respect to $\psi^\dagger_L$ and $\psi^\dagger_R$ in the action, one gets  Dirac equation as
\ba
i\frac{\partial}{c\partial t}\psi_L-mc\psi_L+i\sigma.\nabla\psi_R=0, \nonumber \\
i\frac{\partial}{c\partial t}\psi_R+mc\psi_R+i\sigma.\nabla\psi_L=0.
\ea
As discussed in the literature, one can take the nonrelativistic limit for Dirac equation with the nonrelativistic transformation $\psi=e^{-imc^2t}\begin{pmatrix}
\phi_L\\
\phi_R
\end{pmatrix}$ and gets the Schroedinger equation for the two-spin wave function 
\be
i \partial_t \phi_L(t,x)=-\frac{1}{2m}\nabla^2 \phi_L(t,x).
\ee
Now, similar to the scalar field case, we want to define a proper nonrelativistic transformation which the nonrelativistic limit of the particles and antiparticles appear simultaneously in the action. To this end, we present the spinor field as
\be \label{non-rel-trans-full-dirac}
\psi=A(x)+B^\dagger(x) = e^{-imc^2t}\begin{pmatrix}
    \phi_L\\
    \phi_R
\end{pmatrix}+ e^{imc^2t}\begin{pmatrix}
\Phi^*_L\\
\Phi^*_R
\end{pmatrix}.
\ee
Substituting this nonrelativistic transformation in the Dirac Lagrangian density, we find
\ba
\mathcal{L}=\phi^\dagger_L (i\frac{\partial}{c\partial t}\phi_L+i\sigma.\nabla\phi_R)+\phi^\dagger_R(i\frac{\partial}{c\partial t}\phi_R+2mc\phi_R+i\sigma.\nabla\phi_L)+ \nonumber \\
\Phi^{*\dagger}_L (i\frac{\partial}{c\partial t}\Phi^*_L-2mc\Phi^*_R+i\sigma.\nabla\Phi^*_R)+\Phi^{*\dagger}_R(i\frac{\partial}{c\partial t}\Phi^*_R+i\sigma.\nabla\Phi^*_L)+ Oscillating~ terms.
\ea
Using the Reimann-Lebesgue lemma, we can neglect the oscillating term in the action.
Variation relative to $\phi^\dagger_L,\phi^\dagger_R,\Phi^{*\dagger}_L,\Phi^{*\dagger}_R$ gives four dirac equations for two-component objects.
Since we are interested in getting the nonrelativistic limit of the Lagrangian density for $\phi_L$ and $\Phi_R$ which represent the two components of wave function for the spin-$\frac{1}{2}$ particles and antiparticles , we rewrite the Lagrangian density with the on-shell equations relative to $\phi^\dagger_R,\Phi^{*\dagger}_L$ with applying  nonrelativistic approximation $-i\frac{\partial}{c\partial t}\phi_R \ll 2mc\phi_R+i\sigma.\nabla\phi_L$ and $-i\frac{\partial}{c\partial t}\Phi^*_L\ll -2mc\Phi^*_R+i\sigma.\nabla\Phi^*_R$:

\ba
\mathcal{L}=\phi^\dagger_L (i\frac{\partial}{c\partial t}\phi_L+\dfrac{(\sigma.\nabla)^2}{2mc} \phi_L)+ \nonumber \\
\Phi^{*\dagger}_R(i\frac{\partial}{c\partial t}\Phi^*_R+-\dfrac{(\sigma.\nabla)^2}{2mc} \Phi^*_R).
\ea
As a result of the nonrelativistic transformation \eqref{non-rel-trans-full-dirac}, we get the Schroedinger action for the two components of wave function for particle and antiparticle 
\ba
i\partial_t \phi_L=\dfrac{p^2}{2m} \phi_L, ~i\partial_t\Phi_R=\dfrac{p^2}{2m} \Phi_R
\ea
\\

\section{Conclusion}

The relativistic origin of nonrelativistic theories has faced quite extensively in recent
times \cite{Padmanabhan:2017bll,Guth:2014hsa,Namjoo:2017nia,Banerjee:2018pvs,review} and different subtleties and problems associated with the nonrelativistic limit of a scalar field theory to the Schroedinger theory are discussed. In this paper, we have revisited different cases of the nonrelativistic limit of a real and complex scalar field in the level of the Lagrangian and the equation of motion.
As an important example of the scalar field potential, we studied the effect of a potential like $U(\phi)\propto \phi^4$ which can be attributed to axion dark matter field in this limit. Moreover, a formalism for studying the nonrelativistic limit of antiparticles in the quantum mechanics is developed and another approach for the nonrelativistic limit and its problems have been discussed. \\

In accordance with this article analysis, we conclude with some remarks:
\begin{itemize}
        
    \item In the quantization formulation the low-energy, nonrelativistic limit of relativistic quantum field theory is many-body quantum mechanics.

    \item The nonrelativistic limit of the complex scalar field does not have any relativistic correction and the next order relative to $\frac{1}{c}$ will give the Klein-Gordon equation and action for a complex scalar field. For any complex scalar field potential which is not proportional to $U(\phi, \phi^*)=v(x) \phi \phi^*$, the Schroedinger equation becomes nonlinear.
    
    \item In contrast to the complex scalar field, there are some rapidly oscillating terms which appear in the action of a real scalar field. These terms will be disappeared in the nonrelativistic limit and one gets the Schroedinger action. One can see that in this case there are relativist correction terms for the Schroedinger Lagrangian. Similar to the complex scalar field case, for any potential which is not proportional to $U(\phi)=v(x) \phi^2$ the Schroedinger equation becomes nonlinear.
    
    \item It can be seen that the nonrelativistic limit can be applied to the one particle state of the antiparticles and the Schroedinger equation holds for them. They also can have conserved change or mass density in the quantum mechanics.

    \item  
    The are several key features that disappear when we go the quantum mechanics from field theory by the nonrelativistic limit. In nonrelativistic realm, pair creation is no longer achievable. Furthermore, due to the particles-antiparticle asymmetry of our universe, in a typical condensed matter system or our universe, there won't be anything but particles. One needs a field description in which particles and antiparticles are conserved separately, and whose ground state can be defined as containing many particles but in essence no antiparticles. \\
    
    \item We see that the suitable redefinition of the Dirac field could help us to retain the information of anti-particle in the nonrelativistic limit. Since people have found some signs of nonrelativistic antiprotons in the cosmic ray \cite{antiparticles-observation}, it seems it would be more sensible to use that redefinition to obtain some information about the nonrelativistic antiproton when one studies the cosmic ray theoretically.
        
    \item It was discussed different problems associated with the nonlocal approach for the nonrelativistic limit such as inconsistent initial condition with quantum field theory, causality problem,  and singular kernel problem.
     \item  One can see that the proper nonrelativistic limit of Dirac action gives the Schroedinger equation for the spin-$\frac{1}{2}$ wave function of both  antiparticles and particles. 
    
\end{itemize}

Several directions for future research exist; Given the boundary condition for a field at future or past (time) infinity, one can estimate the different oscillating terms order of magnitude in nonrelativistic limit and finds its relativistic correction. One can also examine nonrelativistic limits of a scalar field with different interaction (potential) terms which can also have gauge symmetry and calculate their relativistic correction.\\

{\bf Acknowledgments:}
\\

We would like to thank M.M. Sheikh-Jabbari,  Ali Akbar Abolhasani, and Mahdiyar Noorbala for useful comments and discussion.\\

\clearpage
\appendix
\section{Riemann-Lebesgue lemma for nonrelativistic limit}
The Riemann-Lebesgue lemma states that:

If $ f $ is  integrable on $ \textbf{R}^d $, that is to say, if the Lebesgue integral of $ |f| $ is finite, then the Fourier transform of $ f $ satisfies
\be
 \hat{f}(z):=\int_{\textbf{R}^d} f(x) \exp(-iz \cdot x)\,dx \rightarrow 0\text{ as } |z|\rightarrow \infty.
 \ee

In the nonrelativistic limit of the action for the real scalar field, there were some osculating terms which appear as the integral
\be
\Gamma=\int_V dt e^{imc^2 t} f(t,x).
\ee
The integral by parts yields

\be
 \int_V dt e^{imc^2 t} f(t,x)= \dfrac{ e^{imc^2 t}}{imc^2 }f(t,x)|_{-\infty}^\infty - \int_v dt \dfrac{ e^{imc^2 t}}{imc^2 } \partial_t f(t,x).
\ee
If we repeat the integration by parts again we get
\be
\int_V dt e^{imc^2 t} f(t,x)= \dfrac{ e^{imc^2 t}}{imc^2 }f(t,x)|_{-\infty}^\infty + \dfrac{ e^{imc^2 t}}{m^2c^4  } \partial_t f(t,x)|_{-\infty}^\infty - \int_v dt \dfrac{ e^{imc^2 t}}{m^2c^4 } \partial_t^2 f(t,x).
\ee
Now we estimate the order of magnitude $\Gamma$ in the nonrelativistic limit $c\gg 1$. In the case that $f(t,x)$  be finite and nonzero at past and future time and  its time derivative be integrable then the leading order of $\Gamma \propto \mathcal{O}(\frac{1}{c^2})$. In the case that $f(t,x)|_{-\infty}^\infty =0$ and its time derivative be finite and nonzero and its second time derivative be integrable then then the leading order of $\Gamma \propto \mathcal{O}(\frac{1}{c^4})$. This algorithm can be extended.\\


\end{document}